\documentstyle[twocolumn,aps,pstricks,graphicx]{revtex}

\def\eps{\epsilon}

\begin{document}
\title{Electric coupling to the magnetic resonance of split ring resonators}

\author{N. Katsarakis, T. Koschny, and M. Kafesaki}
\address{Institute of Electronic Structure and Laser (IESL),
Foundation for Research and Technology Hellas (FORTH), \\
P.O. Box 1527, 71110 Heraklion, Crete, Greece.}

\author{E. N. Economou}
\address{IESL-FORTH, P.O. Box 1527,
71110 Heraklion, Crete, Greece, and Dept. Physics, University of
Crete, Greece}

\author{C.  M. Soukoulis}
\address{IESL-FORTH, and Dept. of Materials Science and
Technology, 71110 Heraklion, Crete, Greece}
\address{Ames Laboratory and Dept. Physics and Astronomy, Iowa State
University$^\dagger$, Ames, Iowa 50011}

\author{\parbox[t]{5.5in}{\small %
We study both theoretically and experimentally the transmission
properties of a lattice of split ring resonators (SRRs) for different
electromagnetic (EM) field polarizations and propagation directions.
We find unexpectedly that the incident electric field {\bf E} couples
to the magnetic resonance of the SRR when the EM waves
propagate perpendicular to the SRR plane and the incident {\bf E} 
is parallel to the gap-bearing sides of the SRR.
This is manifested by a dip in the transmission spectrum.
A simple analytic model is introduced to explain this interesting behavior.
   }}

\maketitle


Metamaterials with a negative index of refraction have attracted
recently great attention due to their fascinating electromagnetic (EM) properties.
It was Veselago that introduced the term ``left-handed substances'' in his
seminal work published in 1968 \cite{Veselago}.
He suggested that in a medium for which the permittivity $\eps$ and
permeability $\mu$ are simultaneously negative, the phase of the
EM waves would propagate in a direction opposite to that
of the EM energy flow.
In this case, the vectors {\bf k}, {\bf E} and {\bf H} form a left-handed set and therefore
Veselago referred to such materials as ``left-handed''.
The interest in Veselago's work was renewed since Pendry {\it et al.\@} proposed an
artificial material consisting of
the so-called split-ring resonators (SRRs) which exhibit a band of negative
$\mu$ values in spite of being made of non-magnetic materials,
and wires which provide the negative $\eps$ behavior \cite{Pendry-1}.
Based on Pendry's suggestion and targeting the original idea of Veselago,
Smith {\it et al.\@} demonstrated in 2000 the realization of the first left-handed
material (LHM) which consisted of an array of SRRs and wires, in
alternating layers \cite{Smith-1}.
Since the original microwave experiment by Smith {\it et al.\@} several composite
metamaterials (CMMs) were fabricated \cite{Smith-2,Ozbay} that exhibited
a pass band in which it was assumed that $\eps$ and $\mu$ are both negative.
This assumption was based on transmission measurements of the wires alone,
the SRRs alone, and the CMMs.
The occurrence of a CMM transmission peak within the stop bands of the SRR
and wire structures was taken as evidence for the appearance of LH behavior.
Further support to this interpretation was provided by the demonstration
that such CMMs exhibit negative refraction of EM waves \cite{Shelby,Parazzoli}.
Moreover, there is a significant amount of numerical work
\cite{MS-1,MS-2,Pacheco,MS-3} in which the transmission and reflection data
are calculated for a finite length of metamaterial.
A retrieval procedure can then be applied to obtain the
effective metamaterial parameters $\eps$ and $\mu$, under the assumption
that it can be treated as homogeneous.
This procedure was applied in Ref.~12 
and confirmed that a medium composed
of SRRs and wires could indeed be characterized by effective $\eps$ and $\mu$
whose real parts were both negative over a finite frequency band,
as was the real part of the refractive index $n$.
However, it was recently shown \cite{KKES,katsan}
that the SRRs exhibit resonant electric response in addition to their
resonant magnetic response.
As a result of this electric response and its interaction with the
electric response of the wires, the effective plasma frequency, $\omega_p'$,
is much lower than the plasma frequency of the wires, $\omega_p$.
An easy to apply criterion was proposed \cite{KKES}
to identify if an experimental transmission peak is left-handed (LH) or
right-handed (RH):
If the closing of the gaps of the SRRs in a given LH structure removes only
a single peak from the T data (in the low frequency regime), this is strong
evidence that the T peak is indeed LH.
This criterion is valuable in experimental studies, where one cannot easily
obtain the effective $\eps$ and $\mu$.
It was applied experimentally and it was found that some T peaks that were
thought to be LH turn out to be RH \cite{katsan}.
It seems that a careful study of the SRR behavior, both electric and magnetic,
is necessary for the design and realization of LH structures.
Marqu\'{e}s {\it et al.\@} considered bianisotropy in SRR structures and developed
an analytical model to evaluate the magnitude of cross-polarization
effects \cite{Marques}.

\begin{figure}
\centerline{\includegraphics[clip,width=7cm]{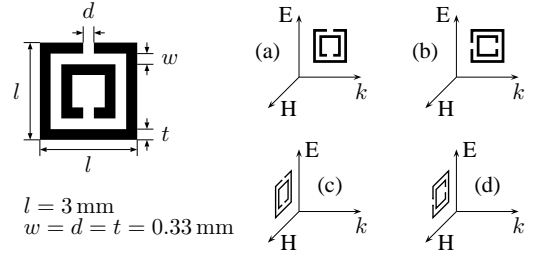}}
\caption{%
Left: The SRR geometry studied. Right: The four studied orientations of the SRR 
with respect to the triad 
{\bf k}, {\bf E}, {\bf H} of the incident EM field.
The two additional orientations, where the SRR are on the 
{\bf H}-{\bf k} plane, produce no electric or magnetic response.
}
\label{fig1}
\end{figure}

In the present paper, we report numerical and experimental results
for the transmission coefficient of a lattice of SRRs alone for
different orientations of the SRR with respect to the external electric
field, {\bf E}, and the direction of propagation.
Incidence is always normal to some face of the orthorhombic unit cell of this
metamaterial, which implies six distinct orientations (Fig.~\ref{fig1}).
It was considered an obvious fact that an incident EM wave excites the
magnetic resonance of the SRR only through its magnetic field;
hence one could conclude that this magnetic resonance appears only 
if the external magnetic field {\bf H}
is perpendicular to the SRR plane, which in turn implies a direction of
propagation parallel to the SRR (Figs.~\ref{fig1}(a), \ref{fig1}(b)).
If {\bf H} is parallel to the SRR (Figs.~\ref{fig1}(c), \ref{fig1}(d))
no coupling to the magnetic resonance was expected.
We show in this paper that this is not always the case.
If the direction of propagation is perpendicular to the SRR plane and
the incident {\bf E} is parallel to the gap-bearing sides of the SRR 
 (Fig.~\ref{fig1}(d)), an electric coupling of the incident
EM wave to the magnetic resonance of the SRR occurs. 
This means that
the electric field excites the resonant oscillation
of the circular current inside the SRR, influencing either the behavior of
$\eps(\omega)$ only (as in Fig. 1(d)) or $\eps(\omega)$ and $\mu(\omega)$
(as in Fig. 1(b)).
Experiments as well as numerical results based on the transfer matrix (TMM),
and on the finite difference time domain (FDTD) method
reveal that for propagation perpendicular to the
SRR plane a dip in the transmission spectrum close to the magnetic resonance
$\omega_m$ of the SRR appears whenever the mirror symmetry of the SRR with
respect to the direction of the electric field is broken by the
gaps of its rings (Fig.~\ref{fig1}(d)).
As we point out below
the possibility of such electric coupling to the magnetic resonance does
also affect the conventional orientations (Figs.~\ref{fig1}(a), \ref{fig1}(b)),
that have the direction of propagation along the SRR plane.
A simple analytic model is given that provides an explanation for the
phenomenon.

For the experimental study, a CMM consisting of SRRs was fabricated
using a conventional printed circuit board process with
$30\,\mu\mathrm{m}$ thick copper patterns on one side of a
$1.6\,\mathrm{mm}$ thick FR-4 dielectric substrate.
The FR-4 board has a dielectric constant of 4.8 and a dissipation
factor of 0.017 at $1.5\,\mathrm{GHz}$.
The design and dimensions of the SRR, which are the same as those of
 Ref.~5, are described in Fig.~1.
The CMM was then constructed by stacking together the SRR structures
in a periodic arrangement.
The unit cell contains one SRR and
has the dimensions $5\,\mathrm{mm}$ (parallel to the cut sides),
$3.63\,\mathrm{mm}$ (parallel to the continuous sides),
and $5.6\,\mathrm{mm}$ (perpendicular to the SRR plane).
The transmission measurements were performed in free space on
a CMM block consisting of 25$\times$25$\times$25 unit cells,
using a Hewlett-Packard 8722 ES network analyzer and microwave
standard-gain horn antennas.

Additionally, numerical simulations using TMM and FDTD method
where performed to understand the couplings to the SRR.
Both methods use a discretized model of the SRR, similar to the one
shown in the inset of Fig.~\ref{fig3}, and periodic boundary conditions
perpendicular to the direction of propagation.
The TMM directly computes the complex transmission and reflection amplitudes
and thus allows us to obtain the effective
medium $\eps(\omega)$ and $\mu(\omega)$ via a retrieval procedure
\cite{SSMS}.
In addition, the FDTD allows to visualize the spatial distribution of
the fields and currents inside the system, 
as a function of time.

\begin{figure}
\centerline{\includegraphics[clip,width=7cm]{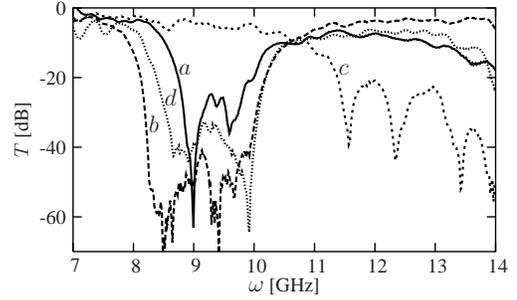}}
\caption{%
Measured transmission spectra of a lattice of SRRs for the four
different orientations shown in Fig.~\ref{fig1}.
}
\label{fig2}
\end{figure}

\begin{figure}
\centerline{\includegraphics[clip,width=7cm]{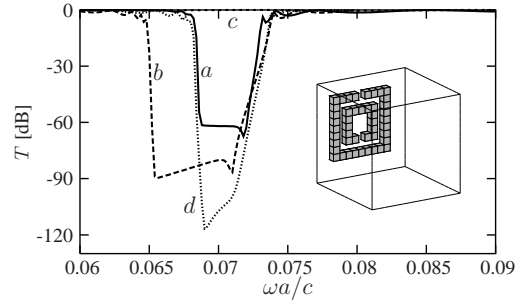}}
\caption{%
Calculated transmission spectra of a lattice of SRRs for the four
different orientations shown in Fig.~\ref{fig1}.
The curve $c$ practically coincides with the axis.
The discretization of one particular SRR is shown in the inset.
}
\label{fig3}
\end{figure}

We considered the four non-trivial orientations of the SRR, which are
shown in Fig.~\ref{fig1}.
Figure \ref{fig2} presents the measured transmission spectra, T, of the CMM.
The continuous line (line $a$) corresponds to the conventional case
shown in Fig.~\ref{fig1}(a), with {\bf H} perpendicular to the SRR plane and
{\bf E} parallel to the symmetry axis of the SRR.
Notice that T exhibits a stop band at
~$8.5$-$10.0\,\mathrm{GHz}$, due to the magnetic resonance.
The dashed line (line $b$) shows T for the orientation of Fig.~\ref{fig1}(b);
here {\bf E} is no longer parallel
to the  symmetry axis of the SRR and thus there is no longer
a mirror symmetry of the combined system of SRR plus EM field.
Notice that now T exhibits a much wider stop band (at 
~$8$-$10.5\,\mathrm{GHz}$), starting at lower frequency.
Very interesting results are obtained by comparing T for the
two cases shown in Figs.~\ref{fig1}(c) and \ref{fig1}(d), for which there is
no coupling to the magnetic field since {\bf H} is parallel to the SRR plane.
For the case \ref{fig1}(c), where {\bf E} is parallel to the symmetry axis,
no structure is observed around the magnetic resonance frequency
(line $c$ in Fig.~\ref{fig2}), as expected.
However in case \ref{fig1}(d), where the 
SRR plus EM field exhibit no mirror symmetry,
a strong stop band in T around $\omega_m$ is observed (line $d$),
similar to that of the conventional case \ref{fig1}(a).
This strongly suggests that the magnetic resonance can be excited by the
electric field provided that there is no mirror symmetry.

\begin{figure}
\centerline{\includegraphics[clip,width=8.cm]{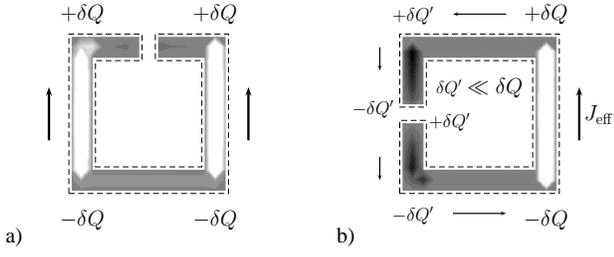}}
\vspace{2mm}
\caption{%
A simple drawing for the polarization in two 
different orientations of a single ring SRR. 
The external electric field points upward. 
Only in case of broken symmetry (b) a circular current will appear 
which excites the magnetic resonance of the SRR.
The interior of the ring shows FDTD data for the 
polarization current component $\mathrm{J}_{\|\mathrm{E}}$ at a fixed time 
for normal incidence (Figs. 1(c),(d)) as a gray scale plot.
}
\label{fig4}
\end{figure}

These observations are in good agreement with the numerical results,
presented in Fig.~\ref{fig3}.
For the propagation perpendicular to the SRR we observe a stop band only
if {\bf E} is parallel to the cut-bearing sides of the SRR and ``sees'' its
asymmetry (line $d$); otherwise we have transparency (line $c$).
At low frequencies,
the SRR can basically be represented only by its outer ring.
As is shown in Fig.~\ref{fig4}, the SRR ring will experience different spatial
distributions of the induced polarization, depending on the relative orientation
of {\bf E} with the SRR gap.
If {\bf E} is parallel to the no gap sides of the SRR its polarization
will be symmetric and the polarization current is only
flowing up and down the sides of the SRR, as shown in Fig.~\ref{fig4}(a).
If the SRR is turned by 90 degree,  shown in Fig.~\ref{fig4}(b), the broken
symmetry leads to a different configuration of surface charges on both sides
of the SRR, connected with a compensating current flowing between the sides.
This current contributes to the circulating current inside the SRR and
hence couples to the magnetic resonance.
We directly observed both types of currents in the FDTD simulations;
as an example, 
the component of the polarization current parallel to the external 
electric field is shown in Fig.~4. 
The retrieval procedure for $\eps$ and $\mu$ indicates that the electric
coupling leads to a resonant electric response in $\eps$ near $\omega_m$.
Also the experimentally observed broadening of the conventional SRR dip for 
the turned SRR was found numerically as well (line $b$ in Fig.~\ref{fig3}).
The reason is the {\em additional} electric coupling which adds
an electric resonant response 
(in $\eps(\omega)$)
directly below the resonant
magnetic response.
Closing the gaps of the SRR \cite{KKES} we observed  both in the
experiment and in the simulations that the dips disappeared.

\smallskip

In summary, we have studied  experimentally and  theoretically the
propagation of EM waves  for different orientations of the SRR.
It is found that the incident electric field couples to the magnetic resonance
of the SRR, provided its direction is such as to break the mirror symmetry.
%
This unexpected electric coupling to the magnetic resonance of the SRR is of
fundamental importance in understanding the refraction properties of
SRRs in the low frequency region of the EM spectrum.
Also this new finding is very important for the design of LHMs
in higher dimensions.

\medskip

\noindent Acknowledgments. Financial support of
EU$\underline{~~}$IST project DALHM, NSF (US-Greece Collaboration),
and DARPA 
are acknowledged. This work was partially supported
by Ames Laboratory.


%
%

\end{document}